\documentclass[fleqn,10pt]{wlscirep}
\usepackage{amsmath,amssymb,bm}
\usepackage{booktabs,multirow,array}
\usepackage{graphicx}
\usepackage{siunitx}
\usepackage{xcolor}
\usepackage{xurl}
\usepackage[normalem]{ulem}
\graphicspath{{./}}

\newcommand{\Rey}{\mathrm{Re}}

\newcommand{\Uinf}{U_\infty}
\newcommand{\avg}[1]{\langle #1\rangle}
\newcommand{\code}[1]{\texttt{#1}}
\newcommand{\nekrs}{nekRS}
\newcommand{\Lz}{L_z}

\title{
Turbulent wakes behind nine bluff-body shapes from direct numerical simulation and particle image velocimetry

} 

\author[1,+,*]{Zhen Zhang}
\author[2,+]{Theo K\"aufer}
\author[2,+]{Louise Ronglan}
\author[2]{Michael S. Triantafyllou}
\author[1]{George Em Karniadakis}
\affil[1]{Division of Applied Mathematics, Brown University, Providence, RI 02912, USA}
\affil[2]{Department of Mechanical Engineering, Massachusetts Institute of Technology, Cambridge, MA 02139, USA}
\affil[+]{these authors contributed equally to this work}
\affil[*]{corresponding author: Zhen Zhang (zhen\_zhang2@brown.edu)}

\begin{abstract}

We present a cross-validated dataset of turbulent wakes behind nine bluff bodies at Reynolds number $\Rey = 10\,000$: a circular cylinder, a 2:1 elliptical cylinder at $0^\circ$ and $90^\circ$, short and long isosceles triangles with the apex upstream or downstream, and a square cylinder at $0^\circ$ and $45^\circ$. Each body was studied by wall-resolved direct numerical simulation (DNS) with  a spectral-element solver and by time-resolved planar particle image velocimetry (PIV) in the MIT Intelligent Towing Tank. The DNS records contain time- and spanwise-averaged velocity, pressure and Reynolds stresses, 24-bin phase-averaged fields keyed to the lift signal, drag and lift histories, and one three-dimensional instantaneous snapshot per body; the PIV records contain the matching time-mean and phase-averaged fields in the same coordinate frame. The DNS set-up is verified against reference cylinder data at $\Rey = 3900$, and DNS
and PIV at $\Rey = 10,000$ are compared field by field. The dataset supports the development and testing of data-driven turbulence closures, reduced-order models, and wake-stability analyses across geometries with distinct separation behavior.

\end{abstract}

\begin{document}
\flushbottom
\maketitle
\thispagestyle{empty}

\section*{Background \& Summary}

% ---- sections/01_background ----
Bluff-body wakes are the canonical test bed for separated turbulent flow: a thin laminar boundary layer separates from the body, rolls up into a K\'arm\'an vortex street, and the near wake mixes coherent shedding with small-scale turbulence. Because both the mean field and the unsteady dynamics are sensitive to the separation mechanism, wakes behind bodies with smooth separation (circular and elliptical cylinders) and fixed separation at sharp edges (squares and triangles) differ markedly in wake width, recirculation length, Strouhal number and Reynolds-stress anisotropy\cite{williamson1996,norberg2003,dong2006}. The circular cylinder at $\Rey = 3900$ has become the reference case for numerical methods\cite{parnaudeau2008,kravchenko2000}, and high-quality experimental and numerical data exist for the square cylinder\cite{lyn1995,trias2015}. What is largely missing, however, is a \emph{single} dataset that covers several shapes at one Reynolds number with identical numerics, identical post-processing, the same experimental facility, and an explicit cross-check between simulation and experiment. While phase-averaged analysis is widely used in vortex shedding literature, fully cross-validated, phase-resolved field datasets are rarely made openly available in this detail.
Such a dataset is needed to train and, more importantly, to test data-driven turbulence closures\cite{duraisamy2019,zhang2026cyl,zhang2026jfm}, physics-informed neural networks\cite{raissi2019}, reduced-order and modal decompositions\cite{taira2017,mendez2019} and generative surrogates, whose value lies precisely in generalization across geometries.

\begin{figure}[htbp]
\centering
\includegraphics[width=0.8\linewidth]{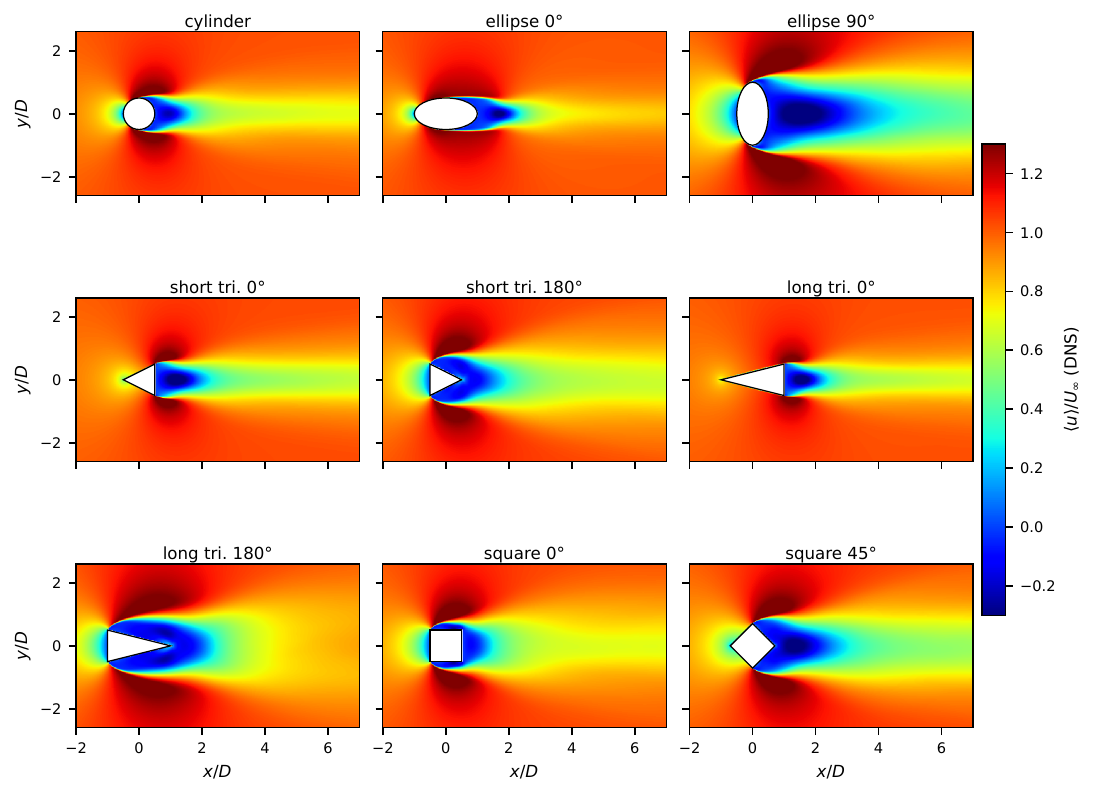}
\caption{\textbf{Overview of the dataset.} Time- and spanwise-averaged streamwise velocity from the DNS for the nine bodies at $\Rey = 10\,000$ in the common coordinate frame.}
\label{fig:overview}
\end{figure}

Here we describe a dataset of nine wakes at $\Rey = \Uinf D/\nu = 10\,000$, where $D$ is the cross-stream reference length (cylinder diameter, square side, triangle base, ellipse minor axis) and $\Uinf$ the free-stream (towing) speed. The bodies (Fig.~\ref{fig:overview}) are a circular cylinder; a 2:1 elliptical cylinder with the major axis aligned with the flow ($0^\circ$) or across it ($90^\circ$); a short ($1D$) and a long ($2D$) isosceles triangle, each with the apex pointing upstream ($0^\circ$) or downstream ($180^\circ$); and a square cylinder face-on ($0^\circ$) and edge-on ($45^\circ$, ``diamond''). Every case was computed by wall-resolved DNS with the spectral-element solver \nekrs{}\cite{fischer2022} on 8 GPUs and measured with time-resolved planar PIV in the MIT Intelligent Towing Tank\cite{itt}. The DNS provides three-dimensional fields that are averaged in time and along the homogeneous spanwise direction, and a phase average conditioned on the lift coefficient that isolates the coherent shedding motion. The PIV provides the independent experimental reference for the mean flow and the Reynolds stresses in the midspan plane, and, through the same phase-averaging logic applied to the measured time series, for the coherent motion.

The dataset was generated for, and has already been used in, work on Reynolds-averaged closures and flow inference with physics-informed neural networks, first for the cylinder\cite{zhang2026cyl} and then across bluff-body shapes\cite{zhang2026jfm}. We release all derived products in self-describing HDF5 files with a common coordinate frame and grid, together with the body geometries, the complete set of DNS--PIV comparison figures and a notebook that reads every product. Figure~\ref{fig:overview} shows the mean flow of the nine bodies; Table~\ref{tab:cases} lists the numerical parameters and the integral quantities of each case.

\section*{Methods}

% ---- sections/02_methods ----
\subsection*{Experiments}
The experiments were performed in the Intelligent Towing Tank (ITT) at MIT\cite{itt}, a \SI{10}{m} long basin of \SI{1}{m}$\times$\SI{1}{m} cross-section in which a gantry robot tows submerged models at a prescribed speed. The measurements come from two campaigns, the first covering seven bodies and the second the square at $0^\circ$ and $45^\circ$; their settings are listed in Table~\ref{tab:pivsetup}. Each body was manufactured with a cross-stream reference dimension $D = \SI{25.4}{mm}$ and towed at the speed of its campaign, giving $\Rey = 10\,000$ at the water temperature and kinematic viscosity of that campaign; the submerged span was \SI{0.75}{m} for the cylinder and \SI{0.723}{m} for the other bodies (aspect ratios of about 30 and 28, blockage ratio 2.5\,\%). The bodies extended through the free surface and carried an end plate at the lower end, and the measurement plane was placed at midspan to minimize end effects. A horizontal light sheet at the model midspan was produced by a \SI{450}{nm} pulsed diode laser (RapidImg RapidPulse B-30) positioned downstream of the field of view, and the flow was seeded with polyamide particles of nominal diameter \SI{25}{\micro m}. In the first campaign the images were recorded by an Emergent Zenith HZ-21000-G-M camera with a Canon \SI{35}{mm} lens mounted in a waterproof housing behind the model, while the second campaign was recorded using a KAYA Iron4510 with with a Canon RF-S \qtyrange[range-phrase = --, range-units = single]{14}{30}{\milli\meter} zoom lens. The calibration gave \SI{0.0903}{mm} per pixel in the first campaign and \SI{0.1589}{mm} per pixel in the second. Vector fields were computed with DaVis~11 (LaVision GmbH) by multi-pass cross-correlation, one pass with $64\times64$-pixel and four passes with $16\times16$-pixel interrogation windows, all at 75\,\% overlap, which gives a vector spacing of four pixels: \SI{0.36}{mm} ($0.014D$) in the first campaign and \SI{0.64}{mm} ($0.025D$) in the second. Outliers were removed by universal outlier detection\cite{westerweel2005}. The records contain 37 shedding cycles for the cylinder, 50 and 16 for the ellipse at $0^\circ$ and $90^\circ$, 34 and 21 for the short triangle at $0^\circ$ and $180^\circ$, 47 and 19 for the long triangle at $0^\circ$ and $180^\circ$, and 26 and 25 for the square at $0^\circ$ and $45^\circ$. Before the statistics were computed, the instantaneous fields of the second campaign were smoothed with the penalized least-squares method of Garcia\cite{garcia2010} in the \code{smoothn} implementation of OpenPIV-Python\cite{openpiv}, each snapshot and velocity component separately and in space only, with the smoothing parameter chosen by generalized cross-validation and without the robust option. All quantities were non-dimensionalized by $D$ and $\Uinf$. 

\begin{table}[htbp]
\centering
\caption{\textbf{PIV campaigns at $\Rey = 10\,000$.} Settings of the two campaigns; the per-body Strouhal numbers and the number of shedding cycles in the phase average are listed in Table~\ref{tab:piv}.}
\label{tab:pivsetup}
\small
% ---- tables/table_piv_setup ----
\begin{tabular}{lp{5.4cm}p{4.8cm}}
\toprule
 & first campaign & second campaign \\
\midrule
Bodies & cylinder, ellipse 0°, ellipse 90°, short tri. 0°, short tri. 180°, long tri. 0°, long tri. 180° & square 0°, square 45° \\
Date & May 2026 & September 2026 \\
Towing speed $\Uinf$ [m/s] & 0.375 & 0.382 \\
Water temperature [$^\circ$C] & 22.2 & 21.4 \\
Kinematic viscosity [$10^{-6}$ m$^2$/s] & 0.953 & 0.970 \\
Scale factor [mm/pixel] & 0.0903 & 0.1589 \\
Acquisition rate [Hz] & 500 & 500 \\
Fields used & 6380 & 6703--6725 \\
Record used [s] ($D/\Uinf$) & 12.8 (188) & 13.4 (202) \\
Shedding cycles in the record & 16--50 & 25--26 \\
Frames per phase bin (24 bins) & 266 & 279--280 \\
Vectors $n_x\times n_y$ & 446$\times$354 & 577$\times$276 \\
Vector spacing [$D$] & 0.0142 & 0.025 \\
Field of view $L_x\times L_y$ [$D$] & 6.3$\times$5.0 & 14.4$\times$6.9 \\
Vector smoothing & none & smoothn\cite{garcia2010} \\
\bottomrule
\end{tabular}

\end{table}

\subsection*{PIV post-processing}
Vectors inside the body and outside the illuminated region are flagged in a mask. Each record is a single tow: the first campaign used 6380 consecutive fields per body (\SI{12.8}{s} at \SI{500}{Hz}, $188\,D/\Uinf$), the second 6703--6725 fields (\SI{13.4}{s}, $202\,D/\Uinf$) after the frames of the gantry deceleration had been removed (Table~\ref{tab:pivsetup}). In space, the field of view measures $L_x\times L_y = 6.3D\times5.0D$ in the first campaign and $14.4D\times6.9D$ in the second (Table~\ref{tab:pivsetup}). Because the coordinate origin of the experiment is defined differently from that of the simulation (Table~\ref{tab:geometry}), the measured fields were translated into the DNS frame; in this common frame the records cover $x/D\in[-0.2, 6.1]$ (cylinder, ellipse at $0^\circ$), $[-0.3, 6.0]$ (ellipse at $90^\circ$), $[0.0, 6.3]$ (short and long triangle at $0^\circ$), $[-0.7, 5.7]$ (short triangle at $180^\circ$), $[-0.8, 5.5]$ (long triangle at $180^\circ$) and $[-2.0, 12.4]$ (square at $0^\circ$ and $45^\circ$), with $|y/D|\le2.4$ for the first campaign and $y/D\in[-2.9, 4.0]$ for the second. All records therefore contain the comparison window $0\le x/D\le5$, $|y/D|\le2$ used in Technical Validation. The time-mean fields $U, V$ and the Reynolds stresses $\avg{u'u'}, \avg{u'v'}, \avg{v'v'}$ were computed over all valid frames of the record. The origin of the PIV coordinates was first placed geometrically and then refined by minimizing the sum of the squared differences of $U$ and $V$ between PIV and the DNS mean field over the window $0.1<x/D<3.5$, $|y/D|<1.5$, with the correction limited to $0.10D$ in each direction. The applied corrections are stored as attributes, and all DNS--PIV differences reported below are evaluated after this registration. The published mean fields were symmetrized about $y=0$ ($U$, $\avg{u'u'}$, $\avg{v'v'}$ even; $V$, $\avg{u'v'}$ odd), keeping the measured value where the mirror point is invalid. For the phase average the cross-stream velocity at the wake point of maximum variance was band-pass filtered between 0.5 and 1.5 times the shedding frequency and its Hilbert phase was used to sort the frames into 24 bins. The phase-averaged Reynolds stresses of the PIV are the covariances of the fluctuations about the bin average, the same definition as in the DNS (below). The PIV phase origin was aligned to the DNS reference by the integer shift of the 24 bins that minimizes the mismatch of the phase-averaged $V$. The Strouhal number of each experiment was obtained from the spectrum of the same probe signal.

\subsection*{Direct numerical simulations}
The incompressible Navier--Stokes equations were solved with \nekrs{} v26\cite{fischer2022}, a GPU spectral-element solver, following the cylinder set-up of Zhang et al.\cite{zhang2026cyl}, in double precision on one node with eight NVIDIA L40S GPUs of the Oscar cluster at Brown University. The domain extends over $y/D\in[-20,20]$, with the streamwise extent and the periodic span $\Lz$ of each case listed in Table~\ref{tab:cases} and the body centered as defined in the geometry files. A uniform velocity $(\Uinf,0,0)$ is imposed at the inlet, a stress-free outflow condition at the outlet, symmetry conditions on the lateral boundaries and no slip on the body. For the cases with the longer domain a fringe (sponge) region $20<x/D<30$ damps the velocity toward the free stream to suppress backflow at the outlet. Body-fitted hexahedral meshes were generated with Gmsh\cite{gmsh}: an O-type layer around the body with the first cell height $\Delta_w$ of Table~\ref{tab:cases} and a geometric growth ratio of 1.3, a refined wake box to $x/D=10$ with near-isotropic cells, and 32 elements across the span. With the polynomial order $N$ of Table~\ref{tab:cases} on Gauss--Lobatto--Legendre points, the meshes have the numbers of elements $N_\mathrm{el}$ and grid points $N_\mathrm{gp}$ listed there, with the first grid point at $y^+<1$ everywhere on the body. Time integration used the second-order backward-difference/extrapolation scheme with characteristics-based sub-cycling of the advection term, an adaptive time step at a fixed target Courant number, and no filtering or subgrid model. Each run was started from rest and advanced for at least $100 D/\Uinf$ until the flow was statistically stationary; the time average and the phase average were then accumulated in two separate runs over the numbers of shedding cycles listed in Table~\ref{tab:cases}.

\begin{table}[htbp]
\centering
\caption{\textbf{Numerical parameters and integral quantities of the DNS at $\Rey = 10\,000$.} Streamwise extent of the domain, spanwise period $\Lz$, first cell height $\Delta_w$ at the wall, $N_\mathrm{el}$ spectral elements, polynomial order $N$, $N_\mathrm{gp}$ grid points (units of $10^8$), number of shedding cycles and duration $T$ (units of $D/\Uinf$) of the time average and of the phase average (two separate runs, both started after the flow had become statistically stationary), mean drag coefficient and r.m.s.\ lift coefficient (frontal height $H$ as reference length) and Strouhal number (reference length $D$).}
\label{tab:cases}
\small
% ---- tables/table_cases ----
\begin{tabular}{lrrrrrrrrrrrrr}
\toprule
Case & $x/D$ & $L_z/D$ & $\Delta_w/D$ & $N_\mathrm{el}$ & $N$ & $N_\mathrm{gp}$ & cycles$_\mathrm{avg}$ & $T_\mathrm{avg}$ & cycles$_\phi$ & $T_\phi$ & $\overline{C_D}$ & $C_L'$ & St \\
\midrule
cylinder & $[-12, 20]$ & $2\pi$ & 0.005 & 153k & 9 & 1.53 & 19 & 99 & 21 & 105 & 1.235 & 0.546 & 0.196 \\
ellipse 0° & $[-12, 30]$ & $1.5\pi$ & 0.005 & 229k & 7 & 1.18 & 110 & 400 & 16 & 58 & 0.608 & 0.115 & 0.275 \\
ellipse 90° & $[-12, 30]$ & $1.5\pi$ & 0.005 & 239k & 7 & 1.23 & 23 & 250 & 17 & 187 & 1.860 & 0.425 & 0.092 \\
short tri. 0° & $[-12, 30]$ & $1.5\pi$ & 0.02 & 223k & 7 & 1.14 & 64 & 300 & 16 & 75 & 1.331 & 0.120 & 0.213 \\
short tri. 180° & $[-12, 30]$ & $1.5\pi$ & 0.02 & 251k & 7 & 1.29 & 27 & 200 & 15 & 123 & 2.398 & 1.178 & 0.136 \\
long tri. 0° & $[-11.5, 20.5]$ & $2\pi$ & 0.005 & 231k & 7 & 1.18 & 35 & 150 & 20 & 81 & 1.079 & 0.207 & 0.236 \\
long tri. 180° & $[-12, 30]$ & $1.5\pi$ & 0.02 & 231k & 7 & 1.18 & 20 & 200 & 12 & 121 & 1.952 & 1.544 & 0.099 \\
square 0° & $[-12, 30]$ & $1.5\pi$ & 0.02 & 242k & 7 & 1.24 & 21 & 160 & 21 & 147 & 2.135 & 1.472 & 0.129 \\
square 45° & $[-12, 30]$ & $1.5\pi$ & 0.02 & 232k & 7 & 1.19 & 19 & 160 & 20 & 161 & 1.848 & 0.515 & 0.120 \\
\bottomrule
\end{tabular}

\end{table}

\subsection*{Statistics computed from the DNS}
\paragraph{Time and spanwise average.}
Within the solver, running time averages of $u_i$, $u_iu_j$, $p$ and $p^2$ are accumulated with the local time step as weight from the start of the averaging window onwards; at every checkpoint the accumulated fields are averaged along the span (which is homogeneous) and written on the unique two-dimensional Gauss--Lobatto--Legendre point set of the mesh. The published mean velocity $U,V$, pressure $P$ and the Reynolds stresses $\avg{u'u'}$, $\avg{u'v'}$, $\avg{v'v'}$, $\avg{w'w'}$ (assuming $\avg{w}=0$) and the pressure variance $\avg{p'p'}$ follow from these moments. Throughout, a prime denotes the fluctuation about the time mean, $u_i = U_i + u_i'$ with $U_i = \avg{u_i}$ and $p = P + p'$, so that $\avg{u_i'u_j'} = \avg{u_iu_j} - U_iU_j$ and $\avg{p'p'} = \avg{p^2} - P^2$. These native point clouds ($2.4$--$3.9\times10^5$ points) are provided together with a linear interpolation on a uniform grid ($x/D\in[-3,8]$, $y/D\in[-3,3]$, spacing $0.02D$) on which the body interior is masked.

\paragraph{Phase average.}
The phase average was accumulated in a second run restarted from the end of the time-averaging run. The instantaneous lift coefficient $C_L(t)$ serves as the phase reference: a rising crossing of $C_L$ through its slow running mean (time constant five shedding periods) starts a new cycle, the phase is $\phi = 2\pi (t - t_0)/T$ with the period $T$ re-measured every cycle, and the spanwise-summed fields $u_i$, $u_iu_j$, $p$, $p^2$ are accumulated into 24 equal bins with the time step as weight. Between 12 and 21 cycles ($T_\phi = 58$--$187\,D/\Uinf$) were accumulated per case (Table~\ref{tab:cases}); the accumulated time per bin is stored so that the slightly shorter first and last partial cycles can be accounted for. The phase average $\tilde u_i(\phi)$ is the average of $u_i$ over all samples of a bin, and a double prime denotes the fluctuation about it, $u_i'' = u_i - \tilde u_i$, so that the phase-averaged Reynolds stresses are $\avg{u_i''u_j''} = \widetilde{u_iu_j} - \tilde u_i\tilde u_j$. In the triple decomposition $u_i = U_i + (\tilde u_i - U_i) + u_i''$ the middle term is the coherent shedding motion: it is excluded from $\avg{u''_iu''_j}$, whereas the time-mean fluctuation $u_i' = (\tilde u_i - U_i) + u_i''$ contains it, so the time-mean stresses $\avg{u'_iu'_j}$ include the coherent contribution.

\paragraph{Force coefficients.}
The pressure and viscous forces on the body were integrated over the full span at every time step and normalized as $C_D = F_x/(\tfrac12\rho\Uinf^2 H \Lz)$ and $C_L = F_y/(\tfrac12\rho\Uinf^2 H\Lz)$ with $H$ the frontal height of the body (Table~\ref{tab:geometry}: $H = D$ except $\sqrt2 D$ for the diamond and $2D$ for the ellipse at $90^\circ$) and $\Lz$ the spanwise period of the case (Table~\ref{tab:cases}); the reference height is stored with each history, so the coefficients can be referred to $D$ by multiplying with $H/D$. Mean drag, lift fluctuation, the Strouhal number from the spectrum of $C_L$, and a 24-bin phase average of $C_D$ and $C_L$ computed with the same crossing detector as the field phase average are provided with each history.

\paragraph{Instantaneous snapshot.}
One checkpoint per case is provided as it was written by the solver, i.e.\ $u,v,w,p$ on the Gauss--Lobatto--Legendre nodes of every spectral element, converted to single precision. Because the meshes are two-dimensional meshes extruded in $z$, the $(x,y)$ coordinates of the $(N+1)^2$ nodes are stored per element and the $z$ levels once per layer, with two index arrays that map each element to its column and layer; the mesh file occupies about $120$\,MB and the field files $0.25$--$1$\,GB per case.

\subsection*{Coordinate conventions}
All published products use the DNS mesh frame: $x$ streamwise, $y$ cross-stream, $z$ spanwise, the origin at the body center for the cylinder, ellipses, square and diamond, and for the triangles at the midpoint of the streamwise extent as listed in Table~\ref{tab:geometry}.
``$0^\circ$'' denotes the sharp apex pointing upstream and ``$180^\circ$'' downstream. The PIV coordinates have been translated into this frame (offsets stored as attributes), so DNS and PIV fields of one case can be overlaid directly.

\begin{table}[htbp]
\centering
\caption{\textbf{Body geometry in the DNS frame} (units of $D$).}
\label{tab:geometry}
\small
\begin{tabular}{lll}
\toprule
Case & definition & frontal height \\
\midrule
cylinder & circle, radius 0.5, center $(0,0)$ & 1 \\
ellipse $0^\circ$ & semi-axes $(1, 0.5)$, center $(0,0)$ & 1 \\
ellipse $90^\circ$ & semi-axes $(0.5, 1)$, center $(0,0)$ & 2 \\
short triangle $0^\circ$ & apex $(-0.5,0)$, base $(0.5,\pm0.5)$ & 1 \\
short triangle $180^\circ$ & base $(-0.5,\pm0.5)$, apex $(0.5,0)$ & 1 \\
long triangle $0^\circ$ & apex $(-1,0)$, base $(1,\pm0.5)$ & 1 \\
long triangle $180^\circ$ & base $(-1,\pm0.5)$, apex $(1,0)$ & 1 \\
square $0^\circ$ & vertices $(\pm0.5,\pm0.5)$ & 1 \\
square $45^\circ$ (diamond) & vertices $(\pm\sqrt2/2, 0)$, $(0,\pm\sqrt2/2)$ & $\sqrt2$ \\
\bottomrule
\end{tabular}
\end{table}

\section*{Data Records}

% ---- sections/03_data_records ----
{\sloppy The dataset is available on the Harvard Dataverse\cite{dataverse} (\url{https://doi.org/10.7910/DVN/4POHU5}) under a CC-BY 4.0 license. It is organized in one folder per case (\code{dns/<case>/}, \code{piv/<case>/}) with the case names \code{cylinder}, \code{ellipse\_0deg}, \code{ellipse\_90deg}, \code{tri\_short\_0deg}, \code{tri\_short\_180deg}, \code{tri\_long\_0deg}, \code{tri\_long\_180deg}, \code{square}, \code{diamond} (the last two measured in the second PIV campaign), plus \code{validation\_re3900/}, \code{validation\_re10k/}, \code{geometry/} and \code{notebooks/}. All field data are HDF5 files whose root attributes record the case, the normalization, the averaging window and the provenance (source file, solver version, applied corrections); every dataset carries a \code{description} attribute. Table~\ref{tab:files} lists the files and their contents.\par}

\begin{table}[htbp]
\centering
\caption{\textbf{Files of each case record.} Shapes refer to $[n_y, n_x] = [301, 551]$ for the DNS grid, $[354, 446]$ for the PIV grid of the May campaign and $[276, 577]$ ($[275, 578]$ for the diamond) for the September campaign, 24 phase bins, and $N_\mathrm{el}$ elements of order $N$ with $n_\mathrm{col}$ element columns and $n_\mathrm{lay}$ layers for the snapshot.}
\label{tab:files}
\small
\resizebox{\textwidth}{!}{
\begin{tabular}{llll}
\toprule
File & group / dataset & shape & content \\
\midrule
\code{dns\_mean.h5} & \code{native/x,y,U,V,P,uu,uv,vv,ww,pp} & $[N_\mathrm{2D}]$ & mean and stresses on the mesh points \\
 & \code{grid/x,y} & $[n_x]$, $[n_y]$ & uniform grid axes \\
 & \code{grid/U,\dots,pp}, \code{body\_mask} & $[n_y,n_x]$ & same on the uniform grid \\
\code{dns\_phase.h5} & \code{phase\_deg}, \code{Tbin}; \code{phase\_bin\_edges\_deg} & $[24]$; $[25]$ & bin centers, accumulated time per bin; bin edges \\
 & \code{native/\dots}, \code{grid/\dots} & $[24,\dots]$ & phase-averaged fields \\
\code{forces.h5} & \code{time,Cd,Cl,Cd\_p,Cd\_v,Cl\_p,Cl\_v,segment} & $[n_t]$ & force history \\
 & \code{phase/phase\_deg,Cd,Cl,Cd\_std,Cl\_std} & $[24]$ & phase-averaged forces \\
\code{snapshot\_mesh.h5} & \code{xy}, \code{z\_levels}, \code{element\_column}, \code{element\_layer} & $[N_\mathrm{el},N{+}1,N{+}1,2]$, $[n_\mathrm{lay},N{+}1]$, $[N_\mathrm{el}]$ & mesh of the snapshot \\
\code{snapshot\_velocity.h5} & \code{u,v,w} & $[N_\mathrm{el},N{+}1,N{+}1,N{+}1]$ & instantaneous velocity \\
\code{snapshot\_pressure.h5} & \code{p} & $[N_\mathrm{el},N{+}1,N{+}1,N{+}1]$ & instantaneous pressure \\
\code{piv\_mean.h5} & \code{x,y,mask,U,V,uu,uv,vv} & $[354,446]$ or $[276,577]$ & PIV mean fields, DNS frame \\
\code{piv\_phase.h5} & \code{phase\_deg}; \code{x,y,mask} & $[24]$; $[n_y,n_x]$ & bin centers; grid and mask as in \code{piv\_mean.h5} \\
 & \code{U,V,uu,uv,vv} & $[24,354,446]$ or $[24,276,577]$ & PIV phase averages \\
 & \code{count}; \code{relL2\_vs\_DNS\_percent\_per\_bin/U,\dots,uv} & $[24]$ & frames per bin; DNS--PIV difference per bin \\
\bottomrule
\end{tabular}}
\end{table}

\paragraph{DNS records.}
\code{dns\_mean.h5} contains the time- and spanwise-averaged $U,V,P,\avg{u'u'},\avg{u'v'},\avg{v'v'},\avg{w'w'},\avg{p'p'}$ in two groups: \code{native/} on the unstructured point set of the mesh (float64, $x,y$ plus eight fields) and \code{grid/} on the uniform grid (float32 arrays of shape $[n_y,n_x] = [301, 551]$, NaN inside the body, with \code{body\_mask}). \code{dns\_phase.h5} contains the 24-bin phase averages of the same eight quantities in the same two groups with an additional leading bin axis, the bin-center phase angles, and the accumulated time per bin. \code{forces.h5} holds the drag and lift histories $C_D, C_L$ and their pressure and viscous parts at every time step of the base and phase-average runs, with the mean drag, lift fluctuation, Strouhal number and the phase-averaged coefficients as attributes and in the \code{phase/} group. \code{snapshot\_mesh.h5}, \code{snapshot\_velocity.h5} and \code{snapshot\_pressure.h5} hold one instantaneous field on the spectral-element mesh: the mesh file gives the $(x,y)$ GLL nodes of each element, the $z$ levels of each layer and the element-to-column/layer maps, the field files give $u,v,w$ and $p$ as float32 arrays of shape $[N_\mathrm{el}, N+1, N+1, N+1]$.

\paragraph{PIV records.}
\code{piv\_mean.h5} contains $U,V,\avg{u'u'},\avg{u'v'},\avg{v'v'}$, the coordinates in the DNS frame and the validity mask on the measurement grid ($354\times446$ for the May campaign, $276\times577$ or $275\times578$ for the September campaign), with the acquisition parameters, the campaign, the number of frames (acquired and, for the September campaign, used), the origin registration (reference origin, applied correction and its limit) and the comparison window as attributes; the September files additionally record the frame window kept after the deceleration was removed and the smoothing applied to the instantaneous fields. \code{piv\_phase.h5} contains the 24-bin phase averages of the same quantities and the number of frames per bin. The raw time-resolved PIV recordings (instantaneous vector fields, about \SI{8}{GB} per body) are not part of the record and are available from the MIT authors on request.

\paragraph{Verification and metadata.}
\code{validation\_re3900/} contains the circular-cylinder verification case at $\Rey=3900$: the mean fields of the verification run in the same format as above, its centerline and profile extractions at $x/D = 1.06, 1.54, 2.02$ and force history, the integral quantities of this and five coarser runs (\code{convergence\_table.csv}), the digitized reference data of Parnaudeau et al.\cite{parnaudeau2008} with their provenance, and a table of literature values. \code{validation\_re10k/} contains the DNS--PIV error tables of Table~\ref{tab:piv}. Its sub-folder \code{spectra/} holds the velocity spectra of Fig.~\ref{fig:spectrum} together with the time series they were computed from: the 41 DNS probe records (40 along the span at $(1.12, \pm0.8)D$ and one at $(5, 0, 2)D$, every time step over $164\,D/\Uinf$) and the PIV vectors at the $3\times3$ points around $(1.12, \pm0.8)D$ and $(5, 0)D$ at \SI{500}{Hz} over the full record. \code{geometry/<case>.json} gives the body vertices or semi-axes in the DNS frame; \code{summary\_cases.csv} collects the parameters and integral quantities of Table~\ref{tab:cases} and \code{dns/summary\_forces.csv} the force statistics of every case; \code{MANIFEST.csv} lists every file with its SHA-256 checksum. \code{notebooks/explore\_dataset.ipynb} is a Jupyter notebook that reads every product of the record and reproduces the comparisons of this Descriptor (Figs.~\ref{fig:re3900}--\ref{fig:errors}). \code{figures/} contains the complete set of DNS--PIV comparison figures generated from the record: the Strouhal-number and error bar charts, the time-mean fields and cross-stream profiles of all five quantities for every body, and the phase-averaged fields of all five quantities at four phases for every body (Figs.~\ref{fig:piv}--\ref{fig:errors} show a selection).

\section*{Technical Validation}

% ---- sections/04_technical_validation ----
\subsection*{Verification of the DNS set-up at Re = 3900}
\begin{figure}[htbp]
\centering
\includegraphics[width=0.85\linewidth]{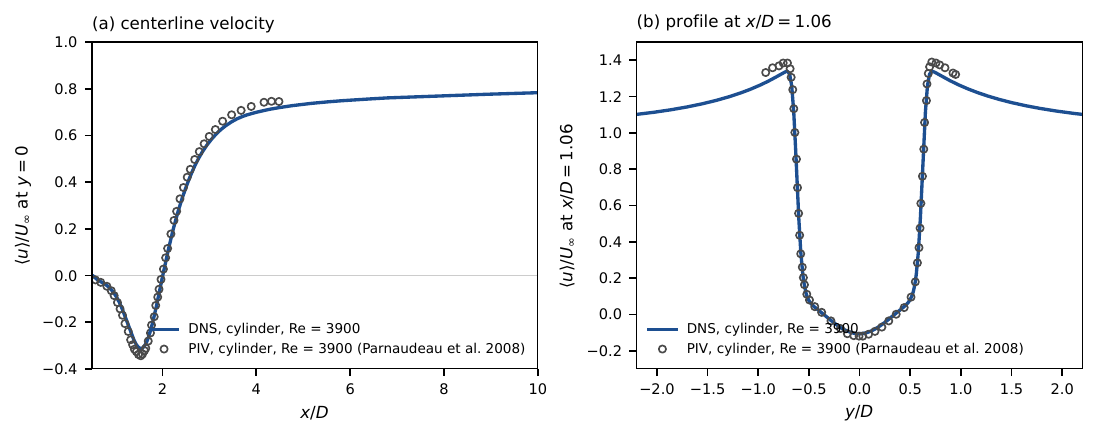}
\caption{\textbf{Verification at $\Rey = 3900$.} Mean streamwise velocity of the DNS verification run compared with the PIV data of Parnaudeau et al.\cite{parnaudeau2008}: (a) on the wake centerline, (b) across the wake at $x/D = 1.06$.}
\label{fig:re3900}
\end{figure}

The numerical set-up (mesh topology, boundary conditions, time integration, averaging) was verified on the circular cylinder at $\Rey = 3900$, for which the PIV measurements of Parnaudeau et al.\cite{parnaudeau2008} and the large-eddy simulations of Kravchenko \& Moin\cite{kravchenko2000} are the accepted references. The verification run uses the same mesh design as the production cases (body-fitted O-layer with a first cell of $0.005D$, wake refinement to $x/D = 10$, 32 element layers over a span of $2\pi D$), with $104\,704$ elements at polynomial order $N = 9$ ($1.0\times10^8$ grid points), an adaptive time step at a target Courant number of one, and statistics accumulated over $400 D/\Uinf$ (about 85 shedding cycles) after a transient of $100 D/\Uinf$. Five coarser runs (fewer elements, $N = 7$, half span, fewer spanwise layers, fixed time step) are included in the record; they over- or under-predict the recirculation length by up to 23\,\%, whereas the three finest runs agree with each other to within 1\,\% in drag and Strouhal number and within 4\,\% in recirculation length. Figure~\ref{fig:re3900} compares the finest run with the PIV data: the centerline velocity (Fig.~\ref{fig:re3900}a) reproduces the measured minimum ($\avg{u}_\mathrm{min}/\Uinf\approx-0.31$ at $x/D\approx1.55$), the recirculation length $L_r/D = 1.51$ (PIV: 1.51) and the recovery, and the cross-stream profile at $x/D = 1.06$ (Fig.~\ref{fig:re3900}b) reproduces the U shape of the near wake and the velocity overshoot at $|y/D|\approx0.7$. The mean drag ($0.99$), Strouhal number ($0.211$), recirculation length ($1.51D$) and base-pressure coefficient ($-0.90$) of this run fall within the spread of the literature values compiled in \code{validation\_re3900/literature\_values.csv}.

\subsection*{Resolution and statistical convergence at Re = 10\,000}
\begin{figure}[htbp]
\centering
\includegraphics[width=0.8\linewidth]{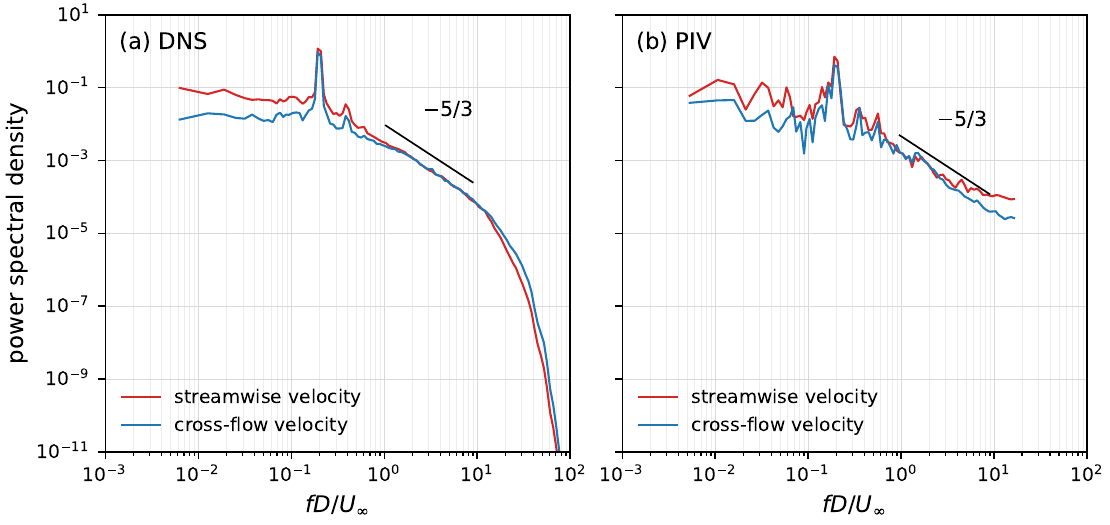}
\caption{\textbf{Velocity spectra in the shear layer of the circular cylinder at $\Rey = 10\,000$.} Power spectral density of the streamwise (red) and cross-stream (blue) velocity at $(x, y) = (1.12, \pm0.8)D$ from (a) the DNS, averaged over 40 probes distributed along the span (record of $159\,D/\Uinf$), and (b) the time-resolved PIV, averaged over the $3\times3$ vectors around the same points (record of $188\,D/\Uinf$, \SI{500}{Hz}). The density is normalised so that its integral over the frequency equals the variance; the straight line marks the $-5/3$ slope. Same location as Fig.~4 of Dong \& Karniadakis\cite{dong2005}.}
\label{fig:spectrum}
\end{figure}

The resolution of the production discretisation was checked with velocity spectra in the separated shear layer of the circular cylinder, following Dong \& Karniadakis\cite{dong2005}. Figure~\ref{fig:spectrum}a shows the power spectral density of the streamwise and cross-stream velocity at $(x, y) = (1.12, \pm0.8)D$, averaged over 40 probes distributed along the span and sampled at every time step over $159\,D/\Uinf$. The spectra peak at the shedding frequency, follow the $-5/3$ law over one decade and then fall by eight orders of magnitude without reaching a noise floor; they match the spectra of Dong \& Karniadakis\cite{dong2005} at the same point, which shows that the dissipative scales of the shear layer are resolved. Figure~\ref{fig:spectrum}b shows the same spectra from the time-resolved PIV record at the same point (\SI{500}{Hz}, $188\,D/\Uinf$, average over the $3\times3$ vectors around $(1.12, \pm0.8)D$): DNS and PIV coincide from the lowest frequencies through the shedding peak and the inertial range up to $fD/\Uinf \approx 5$, beyond which the PIV flattens towards its noise floor at about $10^{-5}$ as the Nyquist frequency ($fD/\Uinf = 17$) is approached.

The time averages span about 20 to 110 shedding cycles (100 to 400 convective units) after a transient of at least 100 units (Table~\ref{tab:cases}). The stationarity of every run was monitored with the force history: the mean drag of the second half of the record differs from that of the first half by less than 5\,\% for seven bodies and by 8\,\% and 12\,\% for the short triangle at $180^\circ$ and the ellipse at $90^\circ$, whose wide wakes exhibit a slow modulation of the shedding amplitude (attributes \code{drift\_Cd\_half}, \code{drift\_Clrms\_half} of \code{forces.h5}). 
The Strouhal number obtained from the lift spectrum agrees with the inverse of the mean cycle length detected by the lift-crossing trigger to within 2\,\% for all cases except the diamond (13\,\%), whose lift signal contains a strong second harmonic. 
For the phase averages, 12 to 21 cycles were accumulated; the two least converged cases (short triangle at $0^\circ$, 16 cycles; long triangle at $180^\circ$, 12 cycles) are flagged in the file attributes and their phase-averaged stresses should be regarded as qualitative.

\begin{figure}[htbp]
\centering
\includegraphics[width=\linewidth]{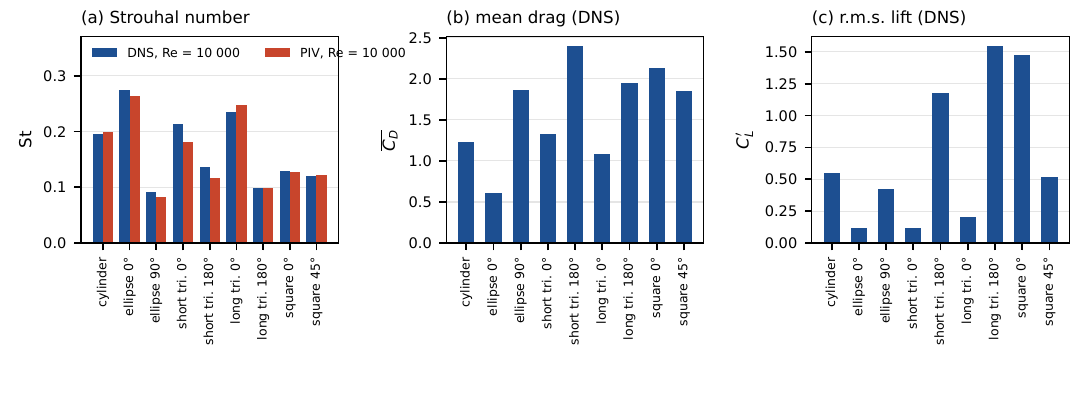}
\caption{\textbf{Integral quantities at $\Rey = 10\,000$.} (a) Strouhal number of the DNS (lift spectrum) and of the PIV (wake-probe spectrum); (b) mean drag coefficient and (c) r.m.s.\ lift coefficient of the DNS, both with the frontal height $H$ as reference length and per unit span.}
\label{fig:integral}
\end{figure}

\subsection*{Discussion of PIV measurement error}
The PIV measurements are affected by measurement errors of different types. The first type of uncertainty is random measurement error, which corresponds to an unbiased scatter around the true value.
A conservative estimate of the uncertainty of a PIV vector is about 0.1 pixel, corresponding to roughly 5 mm/s and 8 mm/s or about 2.5$\%-4\%$ of the speed.
Averaging over multiple measurements reduces the random uncertainty of the mean PIV fields by the square root of the number of snapshots, which lets the random uncertainty diminish. 

Systematic uncertainties, on the other hand, persist through averaging. Most notably is a mismatch between the calibration plane and measurement plane, leading to a systematic over- or underestimation of the
flow velocity and misalignment between the measurement coordinate system and the tow tank gantry coordinate system. Evaluating the magnitude of these errors is not always straightforward, but we estimate the scaling and misalignment error to be well below $5\%$ of the free-stream velocity.
Another important factor is the fluctuation of the tow speed. Our analysis showed that the tow speed fluctuates with a standard deviation of about 0.43$\%$ of the free-stream velocity, with a maximum deviation magnitude of 1.5$\%$.

\subsection*{Cross-validation of DNS and PIV: shedding frequency and mean flow}
\begin{figure}[htbp]
\centering
\includegraphics[width=0.8\linewidth]{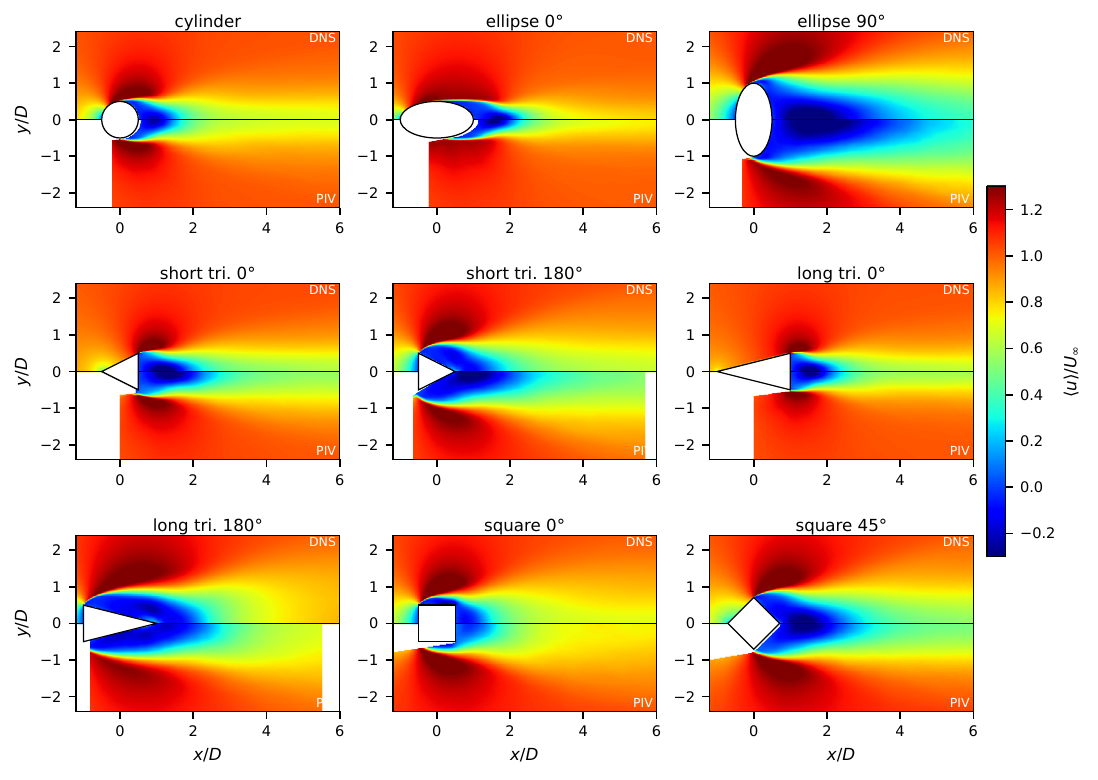}
\caption{\textbf{DNS--PIV comparison of the mean flow at $\Rey = 10\,000$.} Mean streamwise velocity, DNS in the upper and PIV in the lower half-plane, for the nine bodies. The PIV coordinates carry the velocity-registered origin (Methods), so the PIV mask can be displaced from the DNS outline by up to $0.10D$.}
\label{fig:piv}
\end{figure}

\begin{figure}[htbp]
\centering
\includegraphics[width=0.8\linewidth]{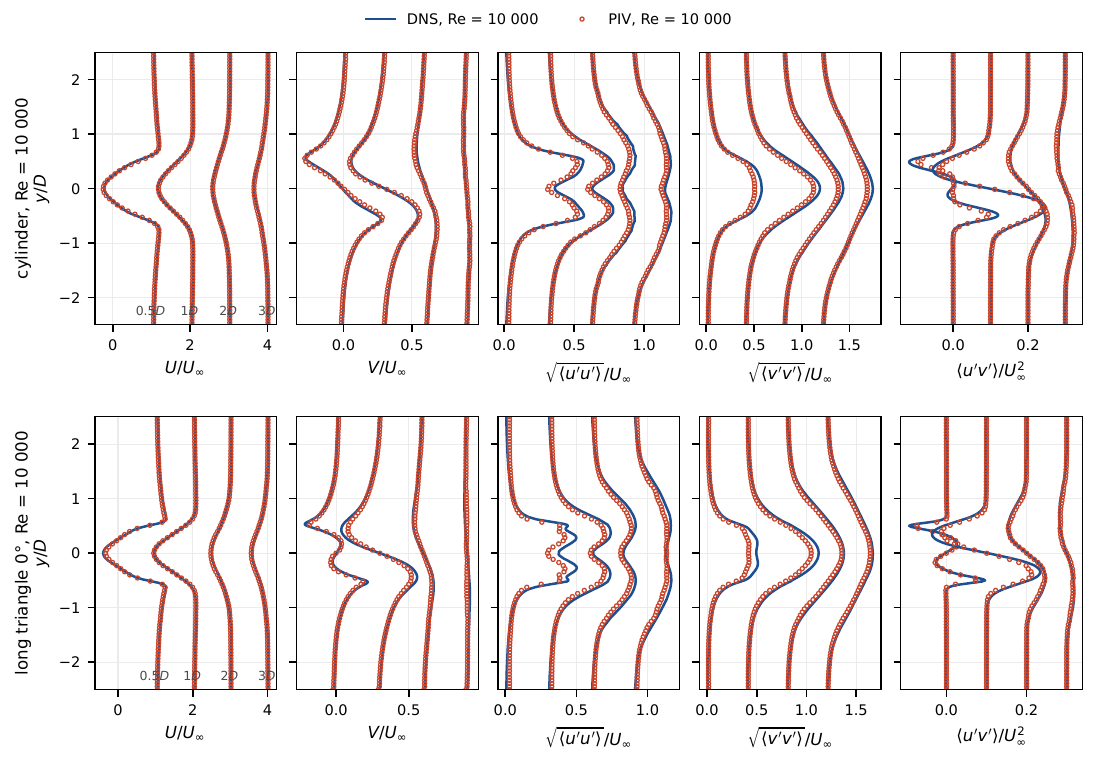}
\caption{\textbf{Time-mean profiles, DNS versus PIV.} Cross-stream profiles of the mean velocities and Reynolds stresses at $x - x_\mathrm{rear} = 0.5$, 1, 2 and $3D$ behind the body (successive stations offset horizontally by 1, 0.3, 0.3, 0.4 and 0.1 for the five quantities) for the circular cylinder (top) and the long triangle at $0^\circ$ (bottom), $\Rey = 10\,000$.}
\label{fig:profiles}
\end{figure}

All DNS--PIV differences quoted below are relative $L_2$ norms over the common comparison window $\Omega = \{0\le x/D\le5,\ |y/D|\le2\}$, the same for all bodies. For a field $f$ the difference is $\epsilon_f = 100\,\|f_\mathrm{DNS}-f_\mathrm{PIV}\|_\Omega/\|f_\mathrm{PIV}\|_\Omega$ with $\|f\|_\Omega^2 = \sum_{\mathbf{x}_k\in\Omega} f(\mathbf{x}_k)^2$, where the sum runs with unit weight over the valid PIV vectors $\mathbf{x}_k$ inside $\Omega$ (vectors inside the body and masked vectors excluded; $9.3$--$9.9\times10^4$ vectors for the first campaign and $3.1\times10^4$ for the second) and the DNS field on the uniform grid is interpolated linearly to these points. The normal stresses enter as their square roots, $f = \sqrt{\avg{u'u'}}$, and for the phase averages the sum runs additionally over the 24 bins, $\|f\|_\Omega^2 = \sum_{b=1}^{24}\sum_{\mathbf{x}_k\in\Omega} f_b(\mathbf{x}_k)^2$. The DNS fields entering $\epsilon_f$ are the time and spanwise averages over the window $T_\mathrm{avg}$ of Table~\ref{tab:cases} (19--110 shedding cycles, 100--400$\,D/\Uinf$) and the phase averages over $T_\phi$ (12--21 cycles, 58--187$\,D/\Uinf$); the PIV fields are the averages over the full records of Table~\ref{tab:pivsetup} (6380 fields, \SI{12.8}{s} or $188\,D/\Uinf$, in the first campaign; 6703--6725 fields, \SI{13.4}{s} or $202\,D/\Uinf$, in the second) and, for the phase averages, over the 266--280 frames of each bin. They are evaluated after the registration of the PIV origin and phase described in Methods, i.e.\ they measure the agreement of the fields up to a translation of at most $0.10D$ and one phase bin. Figure~\ref{fig:integral}a compares the Strouhal numbers of DNS and PIV. They agree to within 6\,\% for six bodies and within 10\,\% for the ellipse at $90^\circ$; the short triangles differ by 15\,\%. Figure~\ref{fig:piv} shows the mean streamwise velocity of all nine cases with the DNS in the upper and the PIV in the lower half-plane: the wake width, the length of the recirculation region and the position of the shear layers coincide for the cylinder, the ellipse at $0^\circ$, the long triangles and the square at $0^\circ$ and $45^\circ$, whereas the ellipse at $90^\circ$ and the short triangles show visible differences that are quantified below. Figure~\ref{fig:profiles} makes the comparison quantitative for the cylinder and the long triangle at $0^\circ$ with profiles of $U$, $V$ and the Reynolds stresses at four stations from $0.5D$ to $3D$ behind the body: the DNS lines and the PIV symbols overlap at all stations, with the only visible departures in the peaks of the shear stress and of $\avg{u'u'}$ at the first station, where the thin separated shear layers are marginally resolved by the PIV interrogation window. Figure~\ref{fig:errors}a and Table~\ref{tab:piv} give the relative $L_2$ difference between DNS and PIV for all bodies, evaluated as defined above. The mean streamwise velocity agrees to within 1.6--20\,\% for all cases and to within 5\,\% for five of them (cylinder, ellipse at $0^\circ$, long triangle at $0^\circ$, square at $0^\circ$ and $45^\circ$); the cross-stream velocity, whose magnitude is small in the mean, agrees to within 10--46\,\%, the r.m.s.\ of the normal stresses to within 9--27\,\% and the shear stress to within 14--47\,\%, the latter being the least converged quantity in both data sets. The largest differences occur for the ellipse at $90^\circ$ and the short triangles, which shed the strongest and most three-dimensional vortices. The recirculation region of the ellipse at $90^\circ$ is $0.7D$ longer in the PIV than in the DNS.

\begin{table}[htbp]
\centering
\caption{\textbf{DNS--PIV differences at $\Rey = 10\,000$.} Relative $L_2$ difference (per cent of the PIV norm) over the comparison window $\Omega$ ($0\le x/D\le5$, $|y/D|\le2$) for the time-mean fields and over $\Omega$ and all 24 phases for the phase-averaged fields, DNS and PIV Strouhal numbers, and number of experimental shedding cycles in the phase average.}
\label{tab:piv}
\resizebox{\textwidth}{!}{
% ---- tables/table_piv_errors ----
\begin{tabular}{lrrrrrrrrrrrr}
\toprule
 & \multicolumn{5}{c}{time mean, rel. $L_2$ [\%]} & & \multicolumn{5}{c}{phase average, rel. $L_2$ [\%]} & \\
Case & $U$ & $V$ & $\sqrt{u'u'}$ & $\sqrt{v'v'}$ & $u'v'$ & St$_\mathrm{DNS}$/St$_\mathrm{PIV}$ & $\tilde U$ & $\tilde V$ & $\sqrt{u''u''}$ & $\sqrt{v''v''}$ & $u''v''$ & cycles$_\mathrm{PIV}$ \\
\midrule
cylinder & 1.6 & 10.4 & 10.4 & 9.1 & 13.6 & 0.196/0.198 & 5.8 & 23.0 & 22.6 & 25.4 & 70.2 & 37 \\
ellipse 0° & 3.6 & 27.8 & 17.6 & 9.7 & 23.9 & 0.275/0.265 & 4.8 & 24.3 & 26.7 & 27.8 & 81.7 & 50 \\
ellipse 90° & 17.7 & 26.2 & 14.5 & 27.0 & 33.3 & 0.092/0.083 & 23.7 & 44.5 & 28.0 & 37.7 & 66.4 & 16 \\
short tri. 0° & 12.8 & 36.4 & 14.0 & 12.0 & 46.4 & 0.213/0.182 & 15.6 & 35.3 & 27.5 & 30.4 & 75.2 & 34 \\
short tri. 180° & 19.8 & 46.2 & 17.1 & 16.5 & 47.3 & 0.136/0.116 & 23.0 & 33.0 & 28.5 & 27.7 & 68.1 & 21 \\
long tri. 0° & 1.8 & 20.8 & 16.7 & 11.8 & 18.9 & 0.236/0.248 & 3.6 & 18.7 & 24.4 & 25.0 & 53.0 & 47 \\
long tri. 180° & 9.1 & 27.0 & 16.8 & 19.6 & 35.8 & 0.099/0.099 & 16.3 & 37.7 & 24.9 & 30.7 & 62.1 & 19 \\
square 0° & 3.7 & 32.5 & 16.8 & 14.4 & 39.4 & 0.129/0.128 & 8.3 & 23.9 & 41.3 & 56.8 & 107.4 & 26 \\
square 45° & 4.4 & 18.7 & 16.6 & 11.6 & 29.6 & 0.120/0.123 & 7.3 & 17.9 & 28.3 & 30.8 & 68.3 & 25 \\
\bottomrule
\end{tabular}

}
\end{table}

\subsection*{Cross-validation of DNS and PIV: phase average}
\begin{figure}[htbp]
\centering
\includegraphics[width=0.8\linewidth]{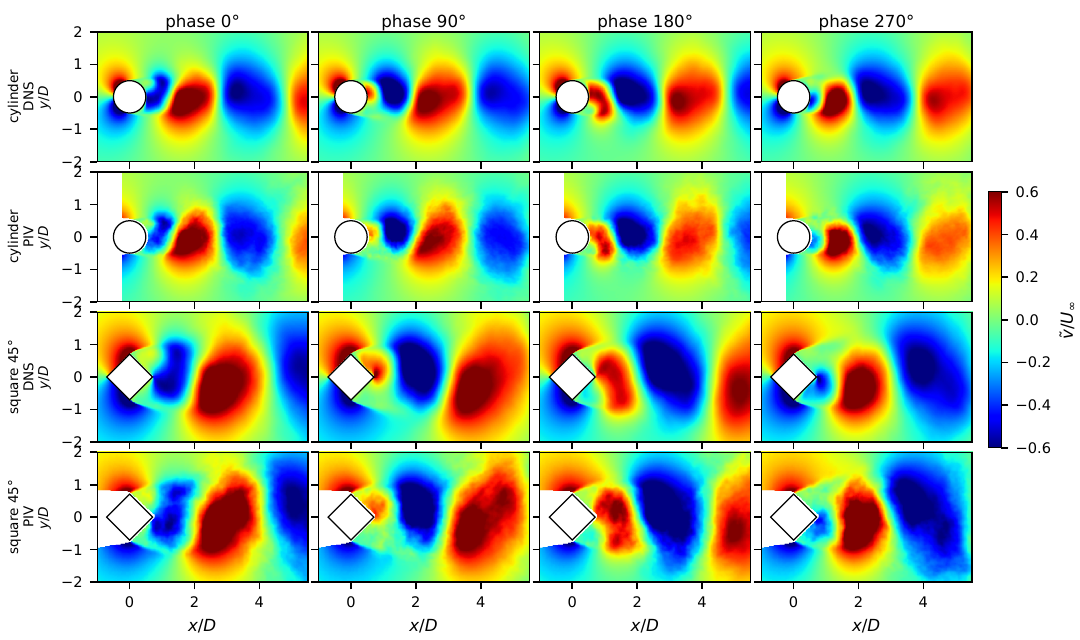}
\caption{\textbf{Phase-averaged comparison.} Phase-averaged cross-stream velocity $\tilde v$ at four phases of the shedding cycle for the cylinder and the diamond, DNS (upper rows) and PIV (lower rows).}
\label{fig:phase}
\end{figure}

\begin{figure}[htbp]
\centering
\includegraphics[width=0.8\linewidth]{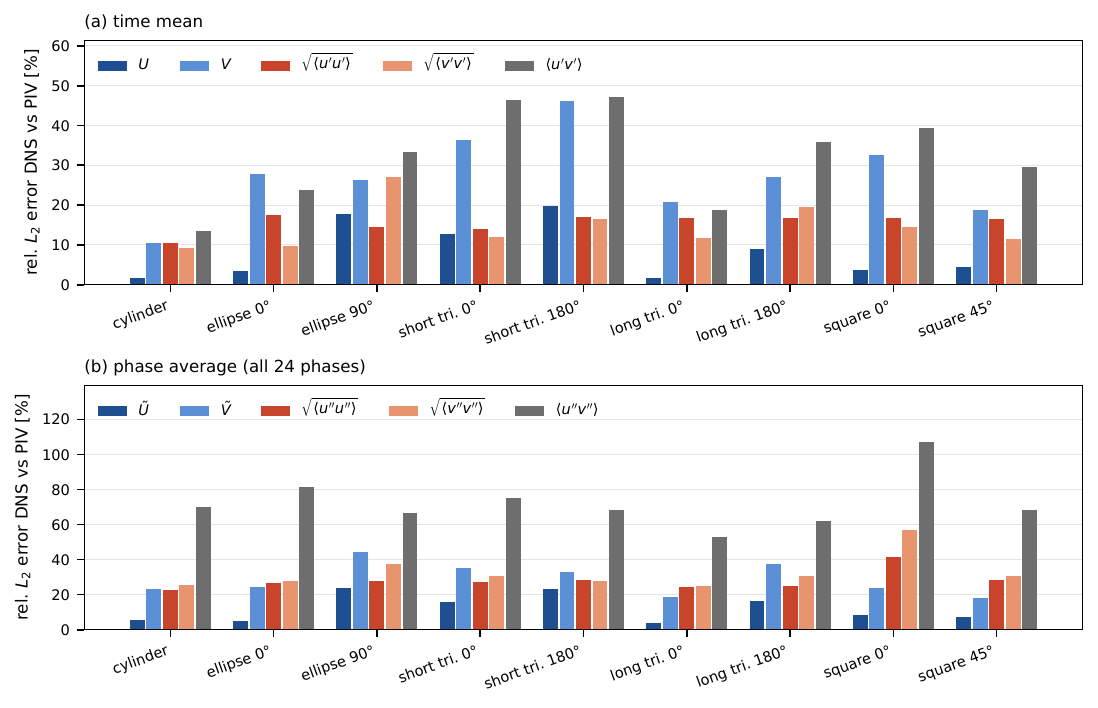}
\caption{\textbf{DNS--PIV differences.} Relative $L_2$ difference between DNS and PIV (per cent of the PIV norm) over the comparison window $\Omega$ ($0\le x/D\le5$, $|y/D|\le2$) for (a) the time-mean velocities and Reynolds stresses and (b) the phase-averaged velocities $\tilde U,\tilde V$ and stresses $\avg{u''u''},\avg{v''v''},\avg{u''v''}$, taken over all 24 phases, for all nine bodies.}
\label{fig:errors}
\end{figure}

Figure~\ref{fig:phase} shows the phase-averaged cross-stream velocity of the cylinder and the diamond at four phases of the shedding cycle in DNS and PIV, and Fig.~\ref{fig:errors}b and Table~\ref{tab:piv} give the relative $L_2$ difference over all 24 phases and the comparison window $\Omega$ for all nine bodies. The position and strength of the shed vortices are captured phase by phase, with $\tilde U$ within 4--24\,\% and $\tilde V$ within 18--45\,\%; the per-bin errors stored in the files vary little over the cycle, showing that the residual is a field-level difference rather than a phase mismatch. The phase-averaged Reynolds stresses differ by 23--38\,\% for the normal components and by 53--82\,\% for the shear stress (41--57\,\% and 107\,\% for the square at $0^\circ$, whose shedding is the least phase-coherent), which reflects the experimental within-bin scatter (about 270 frames per bin) as much as the simulation; the phase-averaged stresses should therefore be regarded as indicative, and the per-bin error curves stored in the files allow users to judge each case.

\subsection*{Known limitations}
(i) Seven DNS cases use a spanwise period of $1.5\pi D$ instead of $2\pi D$; the correlation length of the shedding at $\Rey=10^4$ is well below this value, but the difference should be kept in mind when comparing integral quantities between cases. The mesh and domain study was performed on the circular cylinder at $\Rey=3900$ and was not repeated for the other bodies at $\Rey=10\,000$, for which the resolution evidence is the wall resolution ($y^+<1$), the velocity spectra of Fig.~\ref{fig:spectrum} for the cylinder and the agreement with the PIV. (ii) The phase averages of the short triangle at $0^\circ$ and the long triangle at $180^\circ$ are based on fewer than 16 cycles. (iii) Planar PIV provides neither the spanwise velocity nor the pressure, and the PIV records of the square at $0^\circ$ and $45^\circ$ come from a second campaign with a coarser grid ($0.025D$) and smoothed vector fields, so their shear-layer peaks of the Reynolds stresses are lower than those of the other seven bodies. (iv) The PIV origin has been refined by matching the DNS mean velocity within $0.10D$ in each direction, so the measured body silhouette (the PIV mask) can be displaced from the DNS body outline by up to that amount. The applied shifts are at most $0.03D$ in $y$ and reach the limit in $x$ for five bodies (ellipse at $0^\circ$, both short triangles, long triangle at $180^\circ$, square at $0^\circ$); the geometric origin is recoverable from the stored offsets (\code{origin\_correction\_dx}, \code{origin\_correction\_dy}).

\section*{Usage Notes}

% ---- sections/05_usage_notes ----
All files can be read with any HDF5 library, and the notebook in \code{notebooks/} shows how to load and plot each product. The \code{grid/} groups of the DNS and the PIV files share the coordinate frame, so a DNS field can be sampled at the PIV points with a regular-grid interpolator for point-wise comparison; note that the PIV coordinates carry the velocity-optimized origin, so the PIV mask does not coincide exactly with the DNS body outline (subtract \code{origin\_correction\_dx}, \code{origin\_correction\_dy} to return to the geometric origin); the \code{native/} groups should be used when the full resolution of the near-wall region is required (for example to evaluate wall-normal gradients or to interpolate onto a different mesh). The Reynolds stresses of the time-mean files ($\avg{u'_iu'_j}$, fluctuations about the time mean) include the coherent shedding motion; the stresses of the phase-averaged files ($\avg{u''_iu''_j}$, fluctuations about the phase average) exclude it, so that the triple decomposition of Hussain \& Reynolds\cite{hussain1970} can be formed from the two. The turbulent kinetic energy follows from $k = \tfrac12(\avg{u'u'}+\avg{v'v'}+\avg{w'w'})$ for the DNS; PIV provides only the two in-plane components. Time in the force histories is in units of $D/\Uinf$ and the sampling interval varies with the adaptive time step; the histories should be resampled before spectral analysis. The snapshot fields are periodic in $z$ and, being stored on the solver's own mesh without interpolation, can be used to restart or seed simulations, to test three-dimensional reconstruction methods or as unseen test data for surrogate models; the notebook shows how to assemble the coordinates of an element from the column and layer arrays and how to extract a plane.

\section*{Data availability}
The dataset described in this Descriptor is hosted on the Harvard Dataverse\cite{dataverse} at \url{https://doi.org/10.7910/DVN/4POHU5} under a CC-BY 4.0 license, with the folder structure and file contents given in Data Records (\code{dns/<case>/}, \code{piv/<case>/}, \code{validation\_re3900/}, \code{validation\_re10k/}, \code{geometry/}, \code{notebooks/}, \code{figures/}, \code{summary\_cases.csv}, \code{MANIFEST.csv}). The raw time-resolved PIV recordings are available from the MIT authors on request.

\section*{Code availability}

% ---- sections/06_code_availability ----
The simulations were run with \nekrs{} v26.0.1 (\url{https://github.com/Nek5000/nekRS}) and the PIV vector fields were computed with DaVis~11 (LaVision GmbH) and, for the second campaign, smoothed with OpenPIV-Python\cite{openpiv}. The record includes a Jupyter notebook (\code{notebooks/explore\_dataset.ipynb}, Python~3 with NumPy, SciPy, h5py and Matplotlib) that reads every file of the dataset and reproduces the comparisons shown in this Descriptor; it is the recommended starting point for reusing the data. The \nekrs{} case files (meshes, parameter and user-defined-function files) and the post-processing scripts that produced the record are available from the corresponding author on request.

\section*{Acknowledgements}
The simulations were carried out on the Oscar cluster of the Center for Computation and Visualization (CCV), Brown University.

\section*{Funding}
This research was supported by the Defense Advanced Research Projects Agency (DARPA) under the Automated Prediction Aided by Quantized Simulators (APAQuS) program, Grant No.\ HR00112490526.

\section*{Author contributions statement}
Z.Z. set up and ran the simulations, post-processed the DNS data, performed the DNS--PIV comparison and wrote the manuscript. T.K. and L.R. designed and conducted the experiments and processed the PIV data. M.S.T. and G.E.K. supervised the experimental and computational work, respectively. Z.Z., T.K. and L.R. contributed equally to this work. All authors reviewed the manuscript.

\section*{Competing interests}
The authors declare no competing interests.

\end{document}